\documentclass[twocolumn,showpacs,prb]{revtex4}

\usepackage[cp1251]{inputenc}
\usepackage[russian]{babel}

\usepackage{graphicx}
\usepackage{dcolumn}
\usepackage{bm}

\begin{document}

\title{{{Superconductivity in homologous cuprate series:\\
deep oscillations of pairing Coulomb potential}}}

\author{{V.I. Belyavsky and Yu.V. Kopaev}}

\affiliation{P.N. Lebedev Physical Institute of Russian Academy of
Sciences, Moscow, 119991, Russia}

\begin{abstract}
{Kinematic constraint arising in the case of superconducting
pairing with large momentum results in a cutoff of the screened
Coulomb potential excluding large momentum transfers. This leads
to a pairing potential oscillating in the real space that ensures
a rise of bound singlet pairs. In multilayer cuprates, there is
strong Coulomb interaction between particles composing the pair
not only in the same cuprate layer but in the neighboring layers
as well. In the framework of such a scenario, we explain a
universal dependence of the superconducting transition temperature
on the number of layers in the unit cell observed in homologous
cuprate families.}

\end{abstract}


\maketitle

\noindent {\bf{1}}. All families of superconducting (SC) cuprate
compounds investigated manifest a striking dependence of the
transition temperature $T^{}_c$ on the number $n$ of
${\text{CuO}}^{}_2$ planes in the unit cell. When $n$ increases,
the function $T^{}_c(n)$ grows at first, then, after passing the
maximum at $n=3$, decreases monotonically. \cite{Scott} This
behavior is shown schematically in fig.1 which presents
$T^{}_c(n)$ for a homologous series of mercurocuprates.
\cite{KKCh} One can consider an explanation of such a dependence,
reflecting a fundamental mechanism of SC pairing in the cuprates,
as one of the most important problems of high-$T^{}_c$
superconductors. \cite{Leggett}

A distribution of doped charge turns out to be nonhomogeneous in
multilayer compounds: the inner layers in the unit cell are
underdoped as compared with the outer layers to ensure the minimum
of the electrostatic energy. \cite{Trokiner} Thus, in optimally
(on the average) doped multilayer compound, inner (outer) cuprate
planes are underdoped (overdoped) in comparison with the optimal
doped compound of the family with a single cuprate plane in the
unit cell.

Weak coupling of the neighboring cuprate layers due to coherent
tunnelling of the pairs \cite{CKV} leads to extremely weak initial
rise in the function $T^{}_c(n)$ and its further saturation at
$n>3$. To explain the observed decrease of $T^{}_c(n)$ at $n>3$,
Chakravarty et al.\cite{CKV} took into account a nonhomogeneity of
carrier distribution in the unit cell and also a competition
between the SC and insulating (orbital current $d$~-~symmetry
density wave \cite{CLMN}) orders.

A considerable rise in the SC transition temperature with an
increase of the number of cuprate layers in the unit cell can be
associated with the fact that a spatial scale of the pairing
interaction exceeds the spacing between the neighboring layers. In
contrast to many models based on the extremely localized (in the
real space) SC pairing interaction, such a feature is inherent in
the pairing with large momentum under screened Coulomb repulsion
(${\eta}^{}_K$~-~pairing). \cite{BK}

\begin{figure}
\includegraphics[]{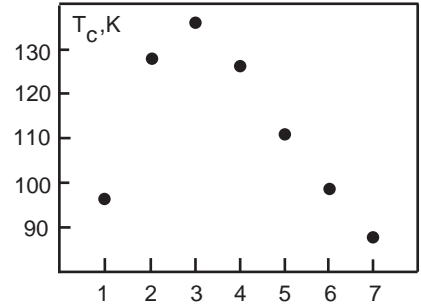}
\caption[*]{{{SC transition temperature ($T^{}_c$) of optimal
doped compound of the family
$HgBa^{}_2Ca^{}_{n-1}Cu^{}_nO^{}_{2n+2+{\delta}}$ vs. a number $n$
of cuprate layers in the unit cell (according to
Ref.[2]).}}}\label{f1}
\end{figure}

\noindent {\bf{2}}. Screening of Coulomb repulsion in classic
electron gas results in the fact that the interaction energy of
two particles at a distance $r$ takes the form
$U(r)=(e^2_{}/r){\exp{(-r/r^{}_0)}}$ corresponding to the Fourier
transform $U(k)=4{\pi}e_{}^2/(k_{}^2+k_0^2)$. Here,
$r^{}_0=k^{-1}_0$ has the meaning of Debye screening length. The
Thomas-Fermi approximation leads to the same expression for the
interaction energy also in the case of degenerate electron gas
with the screening length $r^{}_0 =(4{\pi}e^2_{}n g)_{}^{1/2}$,
where $n$ is the concentration and $g$ is the density of states on
the Fermi level. It should be noted that, when one takes into
account the step-wise electron distribution in the momentum space,
the screened Coulomb potential turns out to be modified
qualitatively, \cite{Ziman}
\begin{equation}\label{1}
U(k)=4{\pi}e_{}^2/[k_{}^2+k_0^2f(k/2k^{}_F)].
\end{equation}
Here, $k^{}_F$ is the Fermi momentum and the Lindhardt function
has the form
\begin{equation}\label{2}
f(x)={\frac{1}{2}} \left (1+{\frac{1-x_{}^2}{4x}}\ln{\left
|{\frac{1+x}{1-x}}\right |}\right ).
\end{equation}
The weak singularity of the potential (\ref{1}) at $k=2k^{}_F$
corresponding to the point of contact of two shifted Fermi
surfaces leads to the damped Friedel oscillations (with the
wavelength ${\pi}/k^{}_F$) of the screened Coulomb potential. The
latter at large $r$ has the form
\begin{equation}\label{3}
U(r)\simeq {\frac{e^2_{}}{2{\pi}}}{\frac{\cos{2k^{}_Fr}}{r^3_{}}}.
\end{equation}
In the case of the nested Fermi surface, the singularity turns out
to be enhanced that can result in a rise of charge or spin density
waves (sctructural phase transition or antiferromagnetic state,
respectively).

Thus, there is a region of the real space where the screened
repulsive Coulomb potential becomes negative. This is sufficient
to ensure SC pairing with non-zero angular momentum of the
relative motion of the pair with zero total momentum. \cite{KL}
However, due to a weakness of the Kohn singularity, the SC
transition temperature turns out to be very low. \cite{KL}

\noindent {\bf{3}}. A rise of a domain $\Xi$ of kinematic
constraint \cite{BK} in the case of SC pairing with large total
momentum ${\bm{K}}$ leads to an enhancement of the singularity of
the pairing potential $U({\bm{k}}-{\bm{k}}_{}^{\prime})$. Indeed,
since in the self-consistency equation that determines the energy
gap parameter ${\Delta}({\bm{k}})$ the momenta of the relative
motion ${\bm{k}}$ and ${\bf{k}}_{}^{\prime}$ are defined (at
$T=0$) only inside a finite region $\Xi$ of the momentum space,
there is a cutoff of far Fourier components of the potential
$U({\bm{k}}-{\bm{k}}_{}^{\prime})$ on the boundary of this region.
It should be noted that, in a general case, the kinetic energy of
the pair at  ${\bm{K}}\neq 0$ vanishes only at some points inside
$\Xi$. Therefore, because of the fact that a logarithmic
singularity in the right-hand side of the self-consistency
equation becomes eliminated, a non-trivial solution to this
equation cannot exist at $U\rightarrow 0$.

We show in fig.2 how the domain of kinematic constraint arises in
the case of isotropic two-dimensional (2D) dispersion. At
${\bm{K}}=0$, the pair state is formed by all one-particle
(electron and hole) states inside the Brillouin zone and the
kinetic energy of the pair vanishes on the whole Fermi surface
(the Fermi contour (FC), in the case of 2D electron system). When
the total momentum ${\bm{K}}$ of the pair increases, the region
$\Xi$ that contributes to the state of the pair decreases down to
zero at $K=2k^{}_F$. At $0<K<2k^{}_F$ the kinetic energy of the
pair vanishes at two points only: these points separate the
regions of the momentum space with electron and hole filling. In
such a case, the logarithmic singularity in the self-consistency
equation is absent.

\begin{figure}
\includegraphics[]{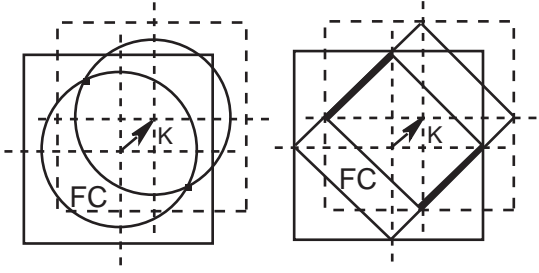}
\caption[*]{{{A rise of the domain of kinematic constraint in the
case of ${\eta}^{}_K$~-~pairing in 2D electron system for
isotropic (left) and nested FC (right); the mirror nesting
condition is fulfilled on the heightened pieces of the
FC.}}}\label{f2}
\end{figure}

The FC with mirror nesting feature, \cite{BK} corresponding to the
coincidence of the energies ${\varepsilon}({\bm{K}}/2+{\bm{k}})$
and ${\varepsilon}({\bm{K}}/2-{\bm{k}})$ of the particles
composing the pair on a finite piece of the FC, results in the
fact that the kinetic energy of the pair vanishes not at the
isolated points but on a finite line inside $\Xi$. It means that,
under mirror nesting condition, the logarithmic singularity in the
self-consistency equation survives and the non-trivial solution
exists at $U\rightarrow 0$. The effective coupling constant in the
exponent of the gap parameter is proportional to the length of the
pieces of the FC on which the mirror nesting condition is
fulfilled,
\begin{equation}\label{4}
{\varepsilon}({\bm{K}}/2+{\bm{k}})={\varepsilon}({\bm{K}}/2-{\bm{k}}).
\end{equation}

The square FC corresponding to the half-filled energy band (fig.2)
manifests apparent mirror nesting for any pair momentum ${\bm{K}}$
directed along one of the diagonals of the Brillouin zone under
the condition that  $K<K^{}_{\pi}$ where
${\bm{K}}^{}_{\pi}=({\pi},{\pi})$. When  ${\bm{K}}$ increases, the
length of the pieces of the FC (on the sides parallel to
${\bm{K}}$) corresponding to the mirror nesting, decreases down to
zero at ${\bm{K}}={\bm{K}}^{}_{\pi}$. Note that at
${\bm{K}}={\bm{K}}^{}_{\pi}$ there is a conventional nesting (for
the opposite sides of the FC),
${\varepsilon}({\bm{p}}+{\bm{K}}^{}_{\pi})=-{\varepsilon}(-{\bm{p}})$,
resulting in the insulating state in the case of the parent
compound. The insulating gap $2{\Delta}^{}_s$ arises at the
position of the FC which becomes a boundary of the magnetic
Brillouin zone of 2D structure with long-range spin
antiferromagnetic (AF) order.

\noindent {\bf{4}}. Thus, the half-filled band of the parent
compound with spin AF order turns out to be split into two
subbabds so that the FC of doped compound appears as small hole
pockets \cite{LNW} near the top of the lower subband. The parts of
the pockets situated in the first magnetic Brillouin zone of the
parent compound (main bands) form the FC with maximum spectral
weight of photoemission. The other parts of the pockets (shadow
bands \cite{KSch} in the second magnetic zone) correspond to the
considerably lower spectral weight decreasing with doping together
with $2{\Delta}^{}_s$. Each pocket manifests perfect mirror
nesting for SC pairs with the momentum ${\bm{K}}^{}_{\pi}$. In
this case, the momentum of one of the particles that compose the
pair belongs to the main band whereas the momentum of the other
particle belongs to the shadow band of the same pocket.

In the case of two pockets situated along the diagonal of 2D
Brillouin zone, there is a perfect conventional nesting with the
same momentum ${\bm{K}}^{}_{\pi}$. The momenta of the particle and
the hole composing the electron-hole pair belong to the different
bands of the pockets: if one of the components of the pair is
related to the main band the other one should be associated with
the shadow band. The FC with both conventional and mirror nesting
features can ensure a competition or coexistence of the SC and
insulating (different from the spin amtiferromagnet) ordered
states.

The ${\eta}^{}_K$~-~pairing channel is efficient under mirror
nesting of the FC. In the case of hole pockets, such a condition
is fulfilled for each of the crystal equivalent momenta
${\bm{K}}^{}_{\pi}= (\pm{\pi},\pm{\pi})$ with their own domains of
kinematic constraint $\Xi$.

To take into account the fact that one of the particles of the
pair belongs to the main band of the hole pocket and the other one
is related to the shadow band, we represent the pairing
interaction in the form
\begin{equation}\label{12}
U^{}_{{\mu}{\mu}_{}^{\prime}}({\bm{k}},{\bm{k}}_{}^{\prime})
={\tilde{v}}^{}_{\mu}({\bm{k}})U({\bm{k}}-{\bm{k}}_{}^{\prime})
{\tilde{u}}^{}_{{\mu}_{}^{\prime}}({\bm{k}}_{}^{\prime}),
\end{equation}
where $\mu$ enumerates the layer in the unit cell. Here,
$U({\bm{k}}-{\bm{k}}_{}^{\prime})$ is the Fourier transform of the
screened Coulomb interaction, ${\tilde{v}}^{}_{\mu}({\bm{k}})$ and
${\tilde{u}}^{}_{{\mu}_{}^{\prime}}({\bm{k}}_{}^{\prime})$ are the
coefficients of the Bogoliubov transformation that diagonalize the
Hamiltonian describing the electron-hole pairing with a rise of
spin AF order; note that
${\tilde{v}}^{}_{\mu}({\bm{k}}+{\bm{K}})\rightarrow
{\tilde{u}}^{}_{\mu}({\bm{k}})$. \cite{CNT} An extension of the
pockets with doping is accompanied with a decrease in the
insulating gap $2{\Delta}^{}_s$ and a strong depression (due to a
deviation from the half-filling) of the spectral weight of the
shadow band $\sim u^{2}_{\mu}({\bm{k}})$. This leads to the
effective restriction (in the momentum space) to the pairing
interaction.

\noindent {\bf{5}}. This can be considered as a reason to replace
the kernel $U({\bm{k}}-{\bm{k}}_{}^{\prime})$ of the pairing
interaction operator by the approximate degenerate kernel
\begin{equation}\label{5}
U({\kappa})=U^{}_0r^d_0(1-{\kappa}_{}^2r_0^2/2),
\end{equation}
presenting two terms of the expansion of the true kernel into the
power series. Here $d=2(3)$ for the two(three)-dimensional system,
$U^{}_0$ and $r^{}_0$ have the meaning of the characteristic
Coulomb energy and the screening length, respectively. The
potential (\ref{5}) manifests itself at small momentum transfer
${\kappa}\equiv|{\bf{k}}-{\bf{k}}_{}^{\prime}|$. Thus, there is an
effective cutoff of the Coulomb repulsion at momenta that are less
than the value $k^{}_c$ which can be considered as a
characteristic length of the domain of kinematic constraint.

At $d=3$, the Fourier transform of the function (\ref{5})
oscillates at large $r$,
\begin{equation}\label{6}
U_{}^{}({\rho})\sim
-{\frac{U^{}_0s^{3}_0}{2{\pi}_{}^2{\rho}^2_{}}} \left
(1-{\frac{s_0^2}{2}}\right ) \cos{{\rho}},
\end{equation}
and has a finite positive value at ${\rho}=0$, where ${\rho}\equiv
k^{}_cr$, $s^{}_0\equiv k^{}_cr^{}_0$. The magnitude of the
oscillation exhibits the maximum at $s^{}_0={\sqrt{6/5}}$ and
decreases with $r$ more slowly as compared to Friedel oscillations
(\ref{3}).

In 2D system ($d=2$), the screened Coulomb potential ensuring the
SC ${\eta}^{}_K$~-~pairing can be estimated (within the framework
of an isotropic model) as
\begin{equation}\label{7}
U_{}^{}({\rho})= {\frac{U^{}_0s^2_0}{2{\pi}{\rho}}} \left
[J^{}_1({\rho}) \left
(1-{\frac{s^2_0}{2}}+{\frac{2s^2_0}{{\rho}^2_{}}} \right )
-{\frac{s^2_0}{{\rho}}}J^{}_0({\rho})\right ],
\end{equation}
where $J^{}_p(x)$ is the Bessel function. The potential (\ref{7})
has a finite value at ${\rho}=0$ and exhibits oscillations slowly
damping at large ${\rho}$:
\begin{equation}\label{8}
U_{}^{}({\rho})\sim {\frac{U^{}_0s^{2}_0}{({\pi}{\rho})_{}^{3/2}}}
\left (1-{\frac{s^2_0}{2}} \right ) \sin{{\rho}}.
\end{equation}
The magnitude of distant oscillations of the potential (\ref{7})
peaks at $s^{}_0=1$.

The potential (\ref{7}) is shown schematically in fig.3 at
$s^{}_0=1$. Note that this potential exhibits more deep (at the
same value of the parameter $s^{}_0$) oscillations as compared to
three-dimensional potential (\ref{6}).

\begin{figure}
\includegraphics[]{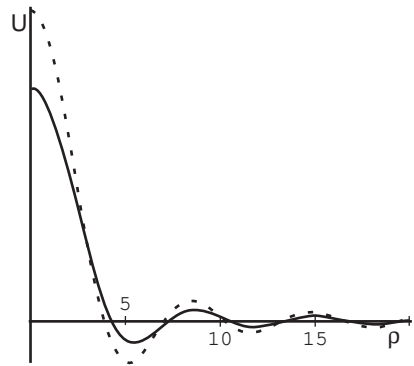}
\caption[*]{{{2D pairing potential (\ref{7}) and step-wise (in the
momentum space) potential (\ref{U}) (in arbitrary units, solid and
dashed lines, respectively).}}}\label{f3}
\end{figure}

Since the cutoff momentum $k^{}_c$, which defines the pairing
interaction (\ref{5}), increases with doping, the dependence of
the kernel magnitude on $s^{}_0$ (initial increase up to the
maximum and the further decrease) can be considered as one of the
reasons resulting in typical of cuprates dependence of the
transition temperature on doping, $T^{}_c(x)$.

The extension of the hole pockets (an increase in the FC length)
with doping promotes a rise in $T^{}_c$ due to the increase of
carrier concentration but, on the other hand, leads to the
decrease in the spectral weight of the shadow band of the FC and,
as a result, the lowering of $T^{}_c$.

\noindent {\bf{6}}. The elimination of all Fourier components of
the screened Coulomb potential that do not belong to the domain
$\Xi$ results in the fact that, in the real space, the pairing
potential exhibits deep damped oscillation with the wavelength
${\pi}/k^{}_c$. Precisely this feature of the screened Coulomb
repulsion becomes apparent in the ${\eta}^{}_K$~-~pairing channel.
In this connection, one can remind the change in Coulomb
repulsion, arising within the framework of the electron-phonon
model of superconductivity, due to the dynamic constraint of the
region of effective attraction in the vicinity of the Fermi
surface, \cite{BTS}
\begin{equation}\label{9}
U^{}_0 \rightarrow U^{}_0/[1+gU^{}_0
\ln{(E^{}_F/{\varepsilon}^{}_D)}],
\end{equation}
where $E^{}_F$ is the Fermi energy, ${\varepsilon}^{}_D$ is the
characteristic Debye energy.

It should be noted that the change of $U(k)$ by a positive
constant $U^{}_0$ inside $\Xi$ leads only to the trivial
(${\Delta}=0$) solution to the self-consistency equation in spite
of the fact that the corresponding potential
\begin{equation}\label{U}
U({\rho})={\frac{U^{}_0s^2_0}{2{\pi}{\rho}}}J^{}_1({\rho})
\end{equation}
oscillates in the real space. Therefore, oscillation of
$U({\rho})$ is a necessary but not yet sufficient condition of the
nontrivial solution existence. Such a solution exists if and only
if the kernel $U({\bm{k}}-{\bm{k}}_{}^{\prime})$ of the pairing
interaction has at least one negative eigenvalue. \cite{BK} It is
obvious that the kernel $U=U^{}_0$ inside $\Xi$ and $U=0$ outside
of $\Xi$ has only one positive eigenvalue.

Degenerate kernel (\ref{5}) that approximates the screened Coulomb
repulsion at small momentum transfers (important in the case of
${\eta}^{}_K$~-~pairing) results in four eigenfunctions (two even
and two odd with respect to the transformation
${\bm{k}}\rightarrow -{\bm{k}}$) even if $r^{}_0$ is arbitrarily
small. A negative eigenvalue corresponds to one of the even
eigenfunctions. A rise of the negative eigenvalue at the
transition from a step-wise kernel to kernel (\ref{5}) can be
compared with the well-known problem of quantum mechanics related
to one-dimensional motion of a particle in asymmetric potential
well: \cite{LLIII} the change in the parameters of the kernel
lowering the degree of asymmetry of the potential (\ref{U}) leads
to splitting off the discrete level from the band of continuous
spectrum.

\noindent {\bf{7}}. Due to the smallness of hopping integrals
between cuprate layers (labelled by ${\mu}=1,2\dots ,n$), the
Fermi surface proves to be opened along the $k^{}_z$~-~axis
corresponding to the $c$~-~axis of the unit cell. The
$k^{}_x,k^{}_y$~-~sections of the Fermi surface corresponding to
the set of layers in the unit cell present the set of the FC's
defined in the sections with different $\mu$. Since the Brillouin
zone size along $k^{}_z$ is $2{\pi}/c$, where $c$ is the size of
the unit cell along the $z$~-~axis, the characteristic separation
between neighboring sections can be estimated as
$k^{}_n=2{\pi}/nc$.

The characteristic size of the domain of kinematic constraint
$k^{}_c\ll{\pi}/a$, where $a$ is the period of the cuprate plane,
determines the wavelength of the pairing potential oscillation in
this plane. It is natural to assume that the momentum $k^{}_0$,
which determines the radius of action of the screened Coulomb
potential in normal to the layers direction, exceeds $k^{}_c$.

This permits of a possibility of SC pairing with the momentum
${\bm{K}}$ not only in the case when the momenta of the particles
${\bm{k}}^{}_{\pm}$, ${\bm{k}}^{\prime}_{\pm}$ (before and after
scattering due to the pairing interaction, respectively) belong to
the same cuprate plane (fig.4a) but also when these momenta are
related to different nearest neighboring planes (fig.4b). Another
possibility of an elevation of $T^{}_c$, considered in Ref.[5], is
connected with the tunnelling of pairs between the neighboring
planes: the momenta ${\bm{k}}^{}_{\pm}$ of particles before
scattering belong to one plane whereas the momenta
${\bf{k}}^{\prime}_{\pm}$ after scattering belong to the
neighboring one (fig.4c).

\begin{figure}
\includegraphics[]{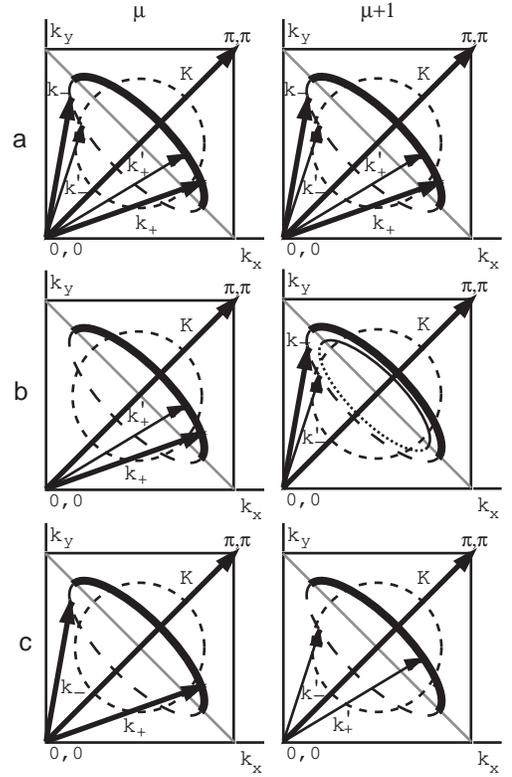}
\caption[*]{SC pairing with large momentum
${\bm{K}}^{}_{\pi}=({\pm{\pi}},{\pm{\pi}})$. {\bf{a:}} The momenta
of particles composing ${\eta}^{}_K$~-~pair before
(${\bm{k}}^{}_+$ and ${\bm{k}}^{}_-$) and after
(${\bm{k}}^{\prime}_+$ and ${\bm{k}}^{\prime}_-$) scattering
belong to the same cuprate plane. {\bf{b:}} One of the particles
with the momenta ${\bm{k}}^{}_+$ and ${\bm{k}}^{\prime}_+$ after
and before scattering, respectively, is in the layer $\mu$, and
the other particle with the corresponding momenta ${\bm{k}}^{}_-$
and ${\bm{k}}^{\prime}_-$) is in the layer ${\mu}+1$. {\bf{c:}}
Both particles of the pair with the momenta ${\bm{k}}^{}_+$ and
${\bm{k}}^{}_-$ before scattering being in the layer $\mu$ pass
(with the momenta ${\bm{k}}^{\prime}_+$ and ${\bm{k}}^{\prime}_-$)
into the layer ${\mu}+1$. The nodal line of the SC order parameter
is shown by dashed line. In fig.4b is shown a change of the FC
(fine solid and dashed lines correspond to the main and shadow
bands, respectively) due to the nonhomogeneous doping of the
layers.}\label{f4}
\end{figure}

The enhancement of efficiency of the ${\eta}^{}_K$~-~pairing
channel leads to the natural explanation of the dependence of
$T^{}_c$ on the number $n$ of cuprate planes in the unit cell
typical of the homologous cuprate series.

\noindent {\bf{8}}. Let us designate the energy gap parameter
relating to the $\mu$-th section of the Brillouin zone by
${\Delta}^{}_{\mu}({\bm{k}})$. In the case of $n$~-~layer
compound, the self-consistency equation can be written in the form
of a system of quasi-linear integral equations
\begin{equation}\label{10}
{\Delta}^{}_{\mu}({\bf{k}})=-{\frac{1}{2}}\sum_{{\mu}_{}^{\prime}=1}^{n}
\sum_{{\bf{k}}_{}^{\prime}}
A^{}_{{\mu}{\mu}_{}^{\prime}}({\bf{k}},{\bf{k}}_{}^{\prime};T)
{\Delta}^{}_{{\mu}_{}^{\prime}}({\bf{k}}_{}^{\prime}),
\end{equation}
where the sum in the right-hand side is over all sections of the
Brillouin zone taking into account the fact that ${\bf{k}}\in
{{\Xi}^{}_{\mu}}$ in each section. The kernel of this system,
which depends on
${\Delta}^{}_{{\mu}_{}^{\prime}}({\bf{k}}_{}^{\prime})$ and
temperature $T$, has the form
\begin{equation}\label{11}
A^{}_{{\mu}{\mu}_{}^{\prime}}({\bf{k}},{\bf{k}}_{}^{\prime};T)=
{\frac{{U^{}_{{\mu}{\mu}_{}^{\prime}}({\bf{k}},{\bf{k}}_{}^{\prime})}}
{E^{}_{{\mu} {\mu}_{}^{\prime}}({\bf{k}}_{}^{\prime})}}
\tanh{\frac{E^{}_{{\mu} {\mu}_{}^{\prime}}({\bf{k}}_{}^{\prime})}
{2T}}.
\end{equation}
Generally speaking, the FC's do not coincide in different sections
of the Brillouin zone due to the different doping levels of the
layers in the unit cell. For the same reason, the quantities
\begin{equation}\label{11'}
{E^{}_{{\mu} {\mu}_{}^{\prime}}}({\bf{k}})
={\sqrt{{\xi}^{2}_{{\mu} {\mu}_{}^{\prime}}({\bf{k}})
+{\Delta}^{2}_{{\mu}_{}^{\prime}}({\bf{k}})}},
\end{equation}
do not coincide as well. Here, $2{\xi}^{}_{{\mu}
{\mu}_{}^{\prime}}({\bf{k}}) ={\varepsilon}({\bf{K}}/2+{\bf{k}})+
{\varepsilon}({\bf{K}}/2-{\bf{k}})$ is the kinetic energy of two
particles in the layers $\mu$ and ${\mu}_{}^{\prime}$,
respectively.

The nontrivial solution to self-consistency equation system
(\ref{11}) under pairing repulsion has to be the function
alternating sign inside the domains $\Xi$ in each section of the
Brillouin zone: the energy gap parameter changes its sign on the
nodal line (circumference) intersecting the FC (fig.4). The fact
that, in the ${\eta}^{}_K$~-~pairing scheme, the SC order
parameter has a nodal line can be associated with the restriction
to double occupation of sites of the cuprate planes taken into
account under the choice of the ground state wave function in
terms of either Gutzwiller \cite{X} or gossamer \cite{Y}
projection operators.

For estimation, it is convenient to approximate
${\Delta}^{}_{\mu}({\bm{k}})$ by step-wise functions which have
the meaning of average values of the function
${\Delta}^{}_{\mu}({\bm{k}})$ in the regions where they are of
constant signs. Thus, ${\Delta}^{}_{\mu}({\bm{k}})$ can be defined
by the constants: \cite{BKT} ${\Delta}^{}_{\mu}({\bm{k}})=
{\Delta}^{}_{\mu}(1)$, if ${\bm{k}}$ belongs to the part of $\Xi$
inside the nodal line, and ${\Delta}^{}_{\mu}({\bf{k}})=
{\Delta}^{}_{\mu}(2)$, if ${\bm{k}}$ is outside this line.

Such an approximation allows us to reduce the kernel
$U({\bm{k}}-{\bm{k}}^{\prime}_{})$ of the pairing interaction to
the degenerate step-wise kernel that can be defined by three
constants \cite{BKT} $U(11)$, $U(22)$, and $U(12)$. Two of the
constants, $U(11)$ and $U(22)$, describe scattering inside and
outside the nodal line, respectively. The third constant, $U(12)$,
describes scattering between the regions of the momentum space
separated by the nodal line. Such a kernel complies with the Suhl
inequality \cite{SMW} $[U(12)]_{}^2 >U(11)U(22)$ ensuring the SC
pairing under repulsion.

The pairing interaction (\ref{12}) can be reduced into a set of
constants
$U^{}_{{\mu}{\mu}_{}^{\prime}}({\alpha}{\alpha}_{}^{\prime})$
where ${\alpha}, {\alpha}_{}^{\prime}=1,2$. The approximation we
use here allows to transform the system of integral equations
(\ref{10}) into the system of transcendental equations,
\begin{equation}\label{13}
{\Delta}^{}_{\mu}({\alpha})=-{\frac{1}{2}}\sum_{{\mu}_{}^{\prime}=1}^{n}
\sum_{{\alpha}_{}^{\prime}=1}^2
U^{}_{{\mu}{\mu}_{}^{\prime}}({\alpha},{\alpha}_{}^{\prime})
f^{}_{{\mu}{\mu}_{}^{\prime}}({\alpha}_{}^{\prime};T)
{\Delta}^{}_{{\mu}_{}^{\prime}}({\alpha}_{}^{\prime}),
\end{equation}
where
\begin{equation}\label{14}
f^{}_{{\mu}{\mu}_{}^{\prime}}({\alpha};T)=
\sum_{{\bf{k}}\in{{\Xi}_{{\mu}_{}^{\prime}}^{{\alpha}}}}
{\frac{\tanh{[E^{}_{{\mu}
{\mu}_{}^{\prime}}({\alpha};{\bf{k}})/2T]} }{E^{}_{{\mu}
{\mu}_{}^{\prime}}({\alpha};{\bf{k}})}},
\end{equation}
${\Xi}_{{\mu}_{}^{\prime}}^{{\alpha}}$ is the part of the domain
of the kinematic constraint inside (${\alpha}=1$) or outside
(${\alpha}=2$) the nodal line,
\begin{equation}\label{14'}
E^{}_{{\mu} {\mu}_{}^{\prime}}({\alpha};{\bf{k}})=
{\sqrt{{\xi}^{2}_{{\mu} {\mu}_{}^{\prime}}({\bf{k}})
+{\Delta}^{2}_{{\mu}_{}^{\prime}}({\alpha})}}.
\end{equation}
The mean-field SC transition temperature is determined by the
condition that $T\rightarrow T^{}_c-0$ when
${\Delta}^{}_{\mu}(\alpha )\rightarrow 0$.

\noindent {\bf{9}}. To analyze equation system (\ref{13}), we use
the simplest approximation \cite{BKT} consistent with the Suhl
inequality, namely,
\begin{equation}\label{S}
U^{}_{{\mu}{\mu}_{}^{\prime}}({\alpha},{\alpha})=0,\;
U^{}_{{\mu}{\mu}}(12)=2w^{}_0,\; U^{}_{{\mu}{\mu}\pm
1}(12)=2w^{}_1.
\end{equation}
Thus, the pairing in the same plane (fig.4a) is described by the
coupling constant $w^{}_0$ whereas another coupling constant,
$w^{}_1$, is related to the pairing in the nearest neighboring
planes (fig.4b,c). Note that, in the case of tunnel mechanism of
the interlayer pairing (fig.4c; such a mechanism is considered by
Chakravarty et al. \cite{CKV}), $w^{}_1\ll w^{}_0$. On the
contrary, in the case when the momenta of the particles before and
after scattering belong to the neighboring sections of the
Brillouin zone (fig.4b), it is natural to think that both
constants, $w^{}_0$ and $w^{}_1$, are of the same exponent.

At $T\rightarrow T^{}_c-0$, parameters (\ref{14}) can be
represented as
\begin{equation}\label{15}
f^{}_{{\mu}{\mu}_{}^{\prime}}({\alpha};T^{}_c)=
g\int\limits_0^{{\varepsilon}^{}_0} {\tanh{\left
({\frac{\xi}{2T^{}_c}}\right )}}{\frac {d{\xi}}{\xi}} =g\ln{\left
({\frac{2{\gamma}{\varepsilon}^{}_0} {{\pi}T^{}_c}}\right )},
\end{equation}
where $g$ is the density of states per one spin,
${\varepsilon}^{}_0$ is the energy scale of the domain of
kinematic constraint ${\Xi}_{{\mu}}^{{\alpha}}$, and
$\ln{\gamma}=0.577$ is the Euler constant. Due to the
logarithmical dependence of the quantities (\ref{14}) on
${\varepsilon}^{}_0$ one can use mutual energy scale for each of
the domains ${\Xi}_{{\mu}}^{{\alpha}}$.

In the case of one-layer compound, equation system (\ref{13})
leads to the typical of the mean-field theory relation between the
transition temperature and the coupling constant,
\begin{equation}\label{16}
T^{}_c(1)=(2{\gamma}{\varepsilon}^{}_0/{\pi})\cdot
\exp{(-1/gw^{}_0)}.
\end{equation}
Within the approximation we use, two components of the energy gap
parameter have equal absolute values and are of opposite sign:
${\Delta}^{}_1(1)=-{\Delta}^{}_1(2)$.

Equation system (\ref{13}) describing two-layer compound has two
solutions for unknown quantity $f\equiv
g\ln{(2{\gamma}{\varepsilon}^{}_0/{\pi}T^{}_c)}$:
$f^{}_{\pm}=(w^{}_0 \pm w^{}_1)_{}^{-1}$. The solution with the
upper sign is related to more high, as compared to $T^{}_c(1)$,
transition temperature,
\begin{equation}\label{17}
T^{}_c(2)=(2{\gamma}{\varepsilon}^{}_0/{\pi})\cdot
\exp{[-1/g(w^{}_0+w^{}_1)]},
\end{equation}
which is determined by the effective coupling constant $w^{\ast}_2
=w^{}_0+w^{}_1$. The sign distribution of the energy gap parameter
is the same in both planes.

As it follows from (\ref{17}), tunnel coupling between cuprate
planes (fig.4c), when $w^{}_1\ll w^{}_0$, cannot ensure a
significant increase in the transition temperature. Therefore, we
can restrict ourselves to the examination of the pairing
mechanisms shown in fig.4a,b. For the sake of simplicity, we also
put $w^{}_1=w^{}_0$. This approximation leads to the doubling of
the effective coupling constant, $w^{\ast}_2 =2w^{}_0$ in the case
$n=2$.

Such an approximation would have led to the tripling of the
coupling constant in the case of three-layer compound. However,
the restriction due to account of the interaction between only the
nearest neighboring layers results in the deceleration of the
growth of the effective coupling constant: the inner layer has two
neighboring layers whereas each of the outer layers has only one
neighboring layer. At $n=3$, this constant turns out to be equal
to $w_3^{\ast}=2.247w^{}_0$. When $n$ increases, the effective
coupling constant manifests a gradual saturation. In particular,
$w_4^{\ast}=2.252w^{}_0$.

The choice of the kernel of the pairing interaction under the
complementary condition
$U^{}_{{\mu}{\mu}_{}^{\prime}}({\alpha}{\alpha})=0$ corresponding
to the symmetry between the filled and vacant parts of $\Xi$ does
not affect a qualitative estimate of the dependence $T^{}_c(n)$.
An important effect of the asymmetry of the domain of kinematic
constraint is a chemical potential shift arising at the transition
into the SC state. This shift appears to be proportional to the
first power of the gap absolute value and may be related to the
problem of high-energy states \cite{Leggett} that become apparent
in optical conductivity as optical sum rule violation.
\cite{Basov}

\noindent {\bf{10}}. The nonhomogeneity of cuprate layer doping in
the unit cell (which is essential in the case $n\geq 3$) results
in the violation of the mirror nesting condition for interlayer
pairing interaction (fig.4b). This happens since the unequal hole
concentration in different layers. Thus, the FC's in the
corresponding sections of the Brillouin zone should be also
different (fig.4b).

In this sense, the difference of carrier concentrations in the
neighboring layers is similar to the exchange field in SC weak
ferromagnets resulting in the Fulde-Ferrel-Larkin-Ovchinnikov
state \cite{FF,LO} with non-zero (small) condensate momentum. For
this reason, one can expect a small difference between the
condensate momentum and ${\bf{K}}^{}_{\pi}$ in the compound with
$n\geq 3$.

A deviation from mirror nesting (in the case of inter-layer
pairing) eliminating the logarithmic singularity in the
self-consistency equation results in the effective decrease in
$T^{}_c$ in the neighboring layers (which, furthermore, differs
from the optimum $T^{}_c$ for the one-layer compound of the
homologous series). Deviations from the mirror nesting and optimum
doping can be considered as the main reasons for a rise in falling
branch of the function $T^{}_c(n)$ at $n\geq 3$. These reasons
should be strengthened by electrostatic effects when $n$
increases.

\noindent {\bf{11}}. It is clear that equation system (\ref{13})
corresponding to the weak-coupling approximation is inadequate for
the evaluation of $T^{}_c$ because the Coulomb coupling constant
$w^{}_0$ cannot most likely be considered as a small quantity.
More realistic estimation should be obtained within the framework
of a renormalization scheme similar to the McMillan's (developed
\cite{McM} for phonon-mediated pairing) approach: $w^{}_0
\rightarrow w^{}_0/(1+w^{}_0)$.

It should be noted in this connection that an increase in the
coupling constant by itself does not result in the considerable
growth of $T^{}_c$. The latter depends essentially on the
preexponential factor ${\varepsilon}^{}_0$ (according to the
phonon-mediated pairing mechanism, ${\varepsilon}^{}_0$ coincides
with a characteristic Debye energy of several hundreds $K$). Under
the ${\eta}^{}_K$~-~pairing, ${\varepsilon}^{}_0$ is determined by
the energy scale of the domain of kinematic constraint. In the
case of cuprates with the energy band of about of one $eV$ it can
exceed $1000\; K$ in spite of relative smallness of this domain.
Thus, within the framework of such a pairing scheme at reasonable
values of the coupling constant $w^{}_0g\leq 1$, there are no
apparent grounds excluding the possibility of the transition
temperature $T^{}_c\approx 300K$ in 2D compounds with mirror
nesting of the FC.


This work was supported in part by the Russian Foundation for
Basic Research (project nos. 05-02-17077, 06-02-17186).

\end{document}